\documentclass[%
preprint,
showpacs,preprintnumbers,
 amsmath,amssymb,
 aps,
 prc,
 longbibliography,
]{revtex4-1}

\usepackage{graphicx}
\usepackage{dcolumn}

\usepackage{bm}
\usepackage{hyperref}

\begin{document}


\title{Theoretical Analysis of Double Differential Cross Section of Proton, Deuteron and Triton for $p+^7$Li Reaction at 14 MeV }


\author{Jiaqi Hu$^{1,2}$}
\author{ Xiaojun Sun$^1$}
\email{sxj0212@gxnu.edu.cn}
\author{Jingshang Zhang$^3$}
\author{Sheng Wang$^2$}
\author{ Yinlu Han$^{1,3}$}
\affiliation{1. College of Physics, Guangxi Normal University, Guilin 541004, People's Republic of China}
\affiliation{2. School of Nuclear Science and Technology, Xi'an Jiaotong University, Xi'an 710049, People's Republic of China}
\affiliation{3. China Institute of Atomic Energy, P. O. Box 275(41), Beijing 102413, People's Republic of China}          



\date{\today}

\begin{abstract}
Based on the statistical theory of light nucleus reactions (STLN), the description of the complicated emission processes of proton and light composite charged particles are further improved through considering the effects of Coulomb barriers both in incident and different outgoing reaction channels. And the analysis of the reaction channels including the sequential and simultaneous emission processes for $p + ^7$Li reaction is performed in detail. So the partial spectra of all of outgoing particles are also obtained for different reaction processes. The calculated double differential cross sections of total outgoing proton, deuteron and triton at $E_p = 14$ MeV agree well with the available experimental data for different outgoing angles. The ENDF-6 formatted data, which includes all of the reaction cross sections, elastic angular distributions, double differential cross sections of nucleon and light composite charged particles for $p + ^7$Li reaction, are also obtained by PUNF code.
\end{abstract}

\pacs{24.10.-i, 25.40.-h, 28.20.Cz}
\keywords{statistical theory of light nucleus reaction; $p + ^7$Li reaction; light composite charged particle; double-differential neutron-production cross section}

\maketitle

\section{\label{sect1}Introduction}
Lithium isotope $^7$Li is a target material nucleus that plays an important role in the fields of nuclear technology and nuclear engineering, such as the compact accelerator-driven neutron source and International Fusion Materials Irradiation Facility (IFMIF) \cite {Garin2011}. The neutron beam produced by compact accelerator-driven neutron sources can be applied to nondestructive testing and medical treatment \cite{M.Paul2015}. For improving the reliability in design of target system for compact accelerator-driven neutron sources, accurate nuclear data of proton induced reactions on $^7$Li are required for Monte Carlo calculation code.  The reaction data of $p + ^7$Li are important due to not only the value of the applications but also the basic research interest in the field of nuclear reaction. There are many open reaction channels at incident proton energy even below $20$ MeV for $p + ^7$Li reaction, so the reaction mechanism is very complex. The cluster separations, such as $^5$He $\rightarrow n+\alpha$, $^5$Li $\rightarrow p+\alpha$ and $^8$Be $\rightarrow \alpha+\alpha$, are involved besides the sequential particle emission processes. The calculation of the nuclear reaction is sensitive to the spin, excited energy and parity of the nuclear levels, especially the elastic and inelatic scattering angular distribution, double-differential cross section for neutron and proton induced nuclear reactions with $1p-$shell light nuclei involved. Therefore, the detailed study on $p + ^7$Li reaction will extend the knowledge of light charged particle induced nuclear reactions with $1p$-shell light nuclei involved, as well as the accurate structural information of some unstable light nuclei. Furthermore, this study will provide abundant proofs to test the reliability of statistical theory of light nucleus reactions (STLN), which has been applied successfully to calculate the double differential cross sections of outgoing neutrons both for neutron and proton induced nuclear reactions with $1p$-shell light nuclei involved \cite{X.J.Sun2017,Zhang2001Li6,Zhang2002Li7,Sun2009Be9,Duan2010Be9,Zhang2003B10,Zhang2003B11,Sun2008kerma,Zhang1999C12,Sun2007C12,Yan2005N14,Zhang2001O16,Duan2005O16,Duan2007,Zhang2015,Yan 5He}.  

The double differential cross sections can give the most detailed information of the emission particles \cite{Trkov2011}. Therefore double differential cross sections of the emitted particles are very important both for the theoretical calculations and the applications. Due to the lack of the appropriate theoretical method for light nucleus reaction, the evaluation or model calculation of outgoing neutron and light charged particles double differential cross section for $p + ^7$Li reaction is not satisfactory. The complete nuclear reaction data for $p + ^7$Li reaction are scarce both for theory and experiment. There are a few data on partial cross sections for $p + ^7$Li reaction, such as $^7$Li($p,p$), $^7$Li($p,n$), $^7$Li($p,d$) and $^7$Li($p,\alpha$) in ENDF/B-VII.$1$ \cite{M.B.Chadwick2011} and ENDF/B-VIII.$0$ \cite{A.D.Carlson2018}. The nuclear reaction data in ENDF/B-VII.1 and ENDF/B-VIII.0 were given based on R-matrix analysis \cite{Philip2005} and fitting to experimental data. In the JENDL-$4.0$/HE \cite{K.Shibata2011}, evaluation was performed by the multi-channel R-matrix fitting the experimental data at incident proton energy below $10$ MeV, and the CCONE code \cite{Iwamoto.O2013} was used to fit ($p,xn$) spectra in incident proton energy range from $10$ to $200$ MeV. The double differential cross sections of some outgoing charged particles for $p + ^7$Li reaction are not given in those three nuclear reaction databases \cite{M.B.Chadwick2011,A.D.Carlson2018,K.Shibata2011}, and cluster separation processes are not also considered in the evaluations.

In addition, $n + ^6$Li reaction \cite{T.Matsumoto2011} and $n + ^7$Li reaction \cite{D.Ichinkhorloo2011} had been calculated with continuum discretized coupled channels (CDCC) method in $2011$ and $2012$, respectively. After that, proton induced reactions on $^{6,7}$Li \cite{H.Guo2013} are also calculated with CCDC method in $2013$. Based on this method, the double differential cross sections of proton and triton from $p + ^7$Li reaction \cite{H.R.Guo2019} were given recently by one of our collaborators. The calculated double differential cross sections of outgoing proton agree well with the experimental data, but the results of outgoing triton are overestimated at large outgoing angles comparing with the measurements. The same method was applied to calculate the double differential cross sections of outgoing neutron for $n+^{6,7}$Li reactions, and the results are underestimated in some low-emission-energy region comparing with the experimental data \cite{H.R.Guo2019}. Moreover, the double differential cross sections of outgoing deuteron are not given in references \cite{T.Matsumoto2011,D.Ichinkhorloo2011,H.Guo2013,H.R.Guo2019}. One of reasons is that the sequential secondary particle emission processes are not considered in CDCC model. Obviously, for high incident proton energies, the effects of secondary particle emission processes between the discret levels and cluster separations in light nuclear reactions are very important. 

The reaction cross sections for $p + ^7$Li reaction have been successfully analyzed by R-matrix theory  in previous studies, however the analysis of energy-angular spectra of the outgoing particles is still an open problem. There are two main problems in calculating double differential cross sections for neutron and proton induced light nucleus reactions. One is the theoretical description of the emission process from a compound nucleus to the discrete levels of the residual nuclei and from the discrete levels of the first residual nuclei to the discrete levels of the secondary residual nucleus with angular momentum and parity conservation through pre-equilibrium process. Another is the recoil effect, which is very important for light nucleus reaction, must be taken into account exactly to keep energy conservation in different reaction processes. Fortunately, these two problems have been solved by STLN \cite{Zhang2015,X.J.Sun2016}. 

Based on the unified Hauser-Feshbach and exciton model \cite{Zhang1993} , which can describe the particle emission processes between the discrete levels with energy, angular momentum, and parity conservation, STLN is proposed to calculate the double differential cross sections of outgoing neutron and light charged particles for neutron induced reactions with the $1p$-shell nuclei involved. STLN has been applied successfully to calculate the double differential cross sections of outgoing neutron for neutron induced reaction on $^6$Li \cite{Zhang2001Li6},$^7$Li \cite{Zhang2002Li7},$^9$Be \cite{Sun2009Be9,Duan2010Be9},$^{10}$B \cite{Zhang2003B10},$^{11}$B \cite{Zhang2003B11},$^{12}$C \cite{Sun2008kerma,Zhang1999C12,Sun2007C12},$^{14}$N \cite{Yan2005N14},$^{16}$O \cite{Zhang2001O16,Duan2005O16} and $^{19}$F \cite{Duan2007}, and the calculated results very well reproduced the measurements. Furthermore, STLN has been improved to apply to the light charged particle induced nuclear reaction with the $1p$-shell nuclei involved. For example, the double differential cross section of outgoing neutron for $p + ^9$Be reaction had been calculated and analyzed in $2016$, and the calculated results successfully reproduced the measurements \cite{X.J.Sun2016}. In this paper, we will further improve STLN, such as analyzing the probable open reaction channels, considering the Coulomb barrier and the emission ratio of the composite particle, to obtain the double differential cross sections of outgoing charged particles for $p + ^7$Li reaction.

In Sec. \ref{sect2}, the theoretical model used in this work is briefly introduced. The reaction channels of $p + ^7$Li reaction below $20$ MeV are analyzed in detail in Sec. \ref{sect3}. The comparisons between calculated results with experimental data and analysis are given in Sec IV, and summary is given in the last section.  

\section{ THEORETICAL MODEL}\label{sect2}
\subsection{ Theoretical Frame}\label{sect2.1}

Since the approach of decribing neutron induced light nucleus reactions with 1$p$-shell light nuclei involved had been proposed in $1999$ \cite{Zhang1999C12}, many experimental data, especially the double differential cross sections, had been analyzed. In dynamics, the angular momentum coupling and parity effect in pre-equilibrium emission process from discrete levels of the first residual nuclei to the discrete levels of the secondary residual nuclei were proposed to accurately keep the angular momentum and parity conservations \cite{Zhang2001O16}, so the double differential cross sections of secondary emitted particles can be calculated by this theoretical model. In kinematics, the recoil effect is strictly taken into account, and the energy balance of particle emission during the different reaction processes could be held accurately. Recently, this approach has been improved to describe proton induced light nucleus reactions with 1$p$-shell light nuclei involved, and to calculate the double differential cross sections of neutron and light charged particles. For illustrating the physical picture, the fundermental formulas are simply given in this paper. The detailed description of STLN can be found in Refs. \cite{Zhang1999C12,Zhang2015,X.J.Sun2016}. 

Based on the unified Hauser-Feshbach and exciton model \cite{Zhang1993}, the cross sections of the first emitted particles from compound nucleus to the discrete energy levels of the first residual nuclei can be expressed as
\begin{eqnarray}\label{eq1}
\sigma_{m_1,k_1}(E_L)=\sum_{j\pi}\sigma_a^{j\pi}(E_L)\{\sum_{n=3}^{n_{max}}P^{j\pi}(n)
\frac{W_{m_1,k_1}^{j\pi}(n,E^*,\varepsilon_{m_1}^c)}{W_T^{j\pi}(n,E^*)}
+Q^{j\pi}(n)\frac{W_{m_1,k_1}^{j\pi}(E^*,\varepsilon_{m_1}^c)}{W_T^{j\pi}(E^*)}\}.   \nonumber\\
\end{eqnarray}
Where $P^{j\pi}(n)$ is the occupation probability of the $n$-th exciton state in the $j\pi$ channel ($j$ and $\pi$ denote the angular momentum and parity in final state, respectively). $P^{j\pi}(n)$  can be obtained by solving the  $j$-dependent exciton master equation under the conservation of angular momentum in pre-equilibrium reaction processes \cite{Zhang1994}. $Q^{j\pi}(n)$ is the occupation probability of the equilibrium state in  $j\pi$  channel. $W_{m_1,k_1}^{j\pi}(n,E^*,\varepsilon_{m_1}^c)$  is emission rate of the first emitted particle $m_1$ at the $n$-th exciton state with outgoing kinetic energy $\varepsilon_{m_1}^c$ in center-of-mass system (CMS), and $W_T^{j\pi}(n,E^*)$ is total emission rate at the $n$-th exciton state. $W_{m_1,k_1}^{j\pi}(E^*,\varepsilon_{m_1}^c)$ is emission rate of the first emitted particle $m_1$ at the equilibrium state with outgoing kinetic energy $\varepsilon_{m_1}^c$ in CMS, and $W_T^{j\pi}(E^*)$ is total emission rate at the equilibrium state. $E^*$ is excited energy of compound nucleus, $\sigma_a^{j\pi}(E_L)$ is absorption cross section in $j\pi$ channel. In Eq. (\ref{eq1}), the first term in the brace denotes the contribution of the pre-equilibrium process, which dominates the light nucleus reactions with 1$p$-shell light nuclei involved. And the second term in the brace denotes the contribution of the equilibrium process. 

The cross section of the secondary outgoing particle from discrete energy level of first residual nucleus to the discrete energy level of the secondary residual nucleus can be expressed as
\begin{eqnarray}\label{eq2}
\sigma_{k_1 \rightarrow k_2}(n, m_1, m_2)=\sigma_{k_1}(n, m_1)\cdot R_{m_2}^{k_1\rightarrow k_2}(E_{k_1}),
\end{eqnarray}
where $\sigma_{k_1}(n, m_1)$  is cross section of the first emitted particle $m_1$ expressed in Eq. (\ref{eq1}), and $R_{m_2}^{k_1\rightarrow k_2}(E_{k_1})$ is the branching ratio of the secondary outgoing particle $m_2$ from energy level $E_{k_1}$ of first residual nucleus $M_1$ to the energy level $E_{k_2}$ of secondary residual nucleus $M_2$.

The formulas (\ref{eq1}) and (\ref{eq2}) describe the particle emission from compound nucleus to discrete energy levels of first residual nuclei, and from the discrete energy levels of the first residual nuclei to the discrete levels of the secondary residual nuclei with angular momentum and parity conservation through the pre-equilibrium and equilibrium reaction process. Our previous researches indicate that the contributions of the total double differential cross sections of outgoing particle for light nucleus reactions are mainly from the pre-equilibrium emission process \cite{X.J.Sun2017,Zhang2001Li6,Zhang2002Li7,Sun2009Be9,Duan2010Be9,Zhang2003B10,Zhang2003B11,Sun2008kerma,Zhang1999C12,Sun2007C12,Yan2005N14,Zhang2001O16,Duan2005O16,Duan2007,Zhang2015,Yan 5He,X.J.Sun2015,X.J.Sun2016}. Only the equilibrium reaction process does not reproduce the double differential cross sections of the light nucleus reactions. 

The linear momentum-dependent exciton state density model \cite{M.B.Chadwick1991} is used to obtain the Legendre expansion coefficients of the first outgoing particle and its residual nucleus. The double differential cross sections of the first outgoing deuteron, triton, $^3$He, and $\alpha$  are calculated with the improved Iwamoto-Harada model \cite{A.Iwamoto1982,J.S.Zhang93,J.S.Zhang1996}, which describes the light composite particle emissions. The representation of the double differential cross sections of secondary outgoing particle had been obtained by the accurate kinematics in Refs.\cite{Zhang2003B10,Zhang1999C12}. And the representation of the double differential cross sections of cluster separation and three-body breakup process can be found in Refs.\cite{Zhang2001Li6,Zhang2002Li7,Zhang2003B11}. Energy conservation is held strictly during nuclear reaction process in laboratory system (LS) for different reaction processes. A new integral formula \cite{X.J.Sun2015}, which is not compiled in any integral tables or any mathematical softwares, had been employed to describe the double differential cross sections of outgoing particles.

According to the Heisenberg’s uncertainty relation, the level widths and energy resolution could be considered in fitting experimental data. The fitting procedure for double differential cross sections of outgoing particles are performed with Gaussian expansion form. And the transformation formulas from CMS to LS have been given in Ref. \cite{Zhang1999C12}. All of the energy level widths is derived from the experimental measurements \cite{Tilley1992,Tilley2002,Tilley2004} as fixed input parameters.

The optical model is very important to calculate the reduced penetration factor, which determines the emission rate of the first emitted particle. The phenomenological spherical optical model potential is employed in the model calculations. The potential parameters of the incident and ejected channels are determined by various reaction cross sections, and the angular distributions of the elastic and inelastic scattering.

\subsection{ Coulomb Barrier}\label{sect2.2}

Since Coulomb barrier has significant effect for open reaction channels of charged particles, it must be reasonably considered in the calculation for incident channel and outgoing channels.
 
Considering the energy-momentum conservation in CMS for outgoing channel, the definitive kinetic energy $\varepsilon _{{m_1}}^c$ of the first emitted particle can be easily derived as
\begin{eqnarray}\label{eq3}
\varepsilon _{{m_1}}^c = \frac{{{M_1}}}{{{M_C}}}\left( {{E^*} - {B_1} - {E_{{k_1}}}} \right).
\end{eqnarray}
Where $M_1$ is mass of the first residual nucleus after emitting the first particle $m_1$, and $M_C$ is mass of compound nucleus. For convenience, $m_1$ and $M_1$ also denote the first outgoing particle and the first residual nucleus, expectively. $E^*$  is excited energy of compound nucleus. $B_1$ is binding energy of the first emitted particle in compound nucleus. $E_{k_1}$ is excited energy of the $k$-th discrete level of the first residual nucleus.

Considering the energy-momentum conservation in CMS for incident channel, the excited energy of compound nucleus can be expressed as
\begin{eqnarray}\label{eq4}
{E^*} = \frac{{{M_T}}}{{{M_C}}}{E_p} + {B_p},
\end{eqnarray}
where $M_T$ is mass of target nucleus. $E_p$ is kinetic energy of incident particle. $B_p$ is binding energy of incident particle in compound nucleus. According to Eqs. (\ref{eq3}) and (\ref{eq4}), the threshold energy  $E_{th}$ can be calculated.

 Due to effect of Coulomb barrier \cite{G.R.Satchler1991,Peter W1970}, the kinetic energy of first outgoing charged particle must be higher than the Coulomb barrier $V_{Coul}$  , namely $ \varepsilon _{{m_1}}^c  > {V_{Coul}}$. According to the assumption of the spherical nucleus \cite{Zhang2015}, the Coulomb barrier ${V_{Coul}}$  can be approximatively expressed as
\begin{eqnarray}\label{eq5}
{V_{Coul}} = \frac{{{e^2}{Z_{{M_1}}}{Z_{{m_1}}}}}{{{r_C}(A_{{M_1}}^{\frac{1}{3}} + A_{{m_1}}^{\frac{1}{3}})}},
\end{eqnarray}
where $Z_{M_1}$ and $Z_{m_1}$ is charge number of residual nucleus and first outgoing charged particle, respectively. $r_C(=1.2\sim1.5$fm) is charge radii parameter. For proton, deuteron, triton, $^3$He, $\alpha $ and $^5$He, their charge radii $r_C$A$^\frac{1}{3}$ will be substituted by the measurements compiled in Ref. \cite{I.Angeli2013}. 

Therefore, the incident energy $E_p$ must meet Eq. (\ref{eq6}) to open reaction channels, i.e.
 \begin{eqnarray}\label{eq6}
{E_p} > \frac{{{M_C}}}{{{M_T}}}(\frac{{{M_C}}}{{{M_1}}}{V_{Coul}} + {E_{{k_1}}} + {B_1} - {B_p}).
\end{eqnarray}
Obviously, the Coulomb barrier can affect significantly the open reaction channels. It is necessary that the reduced penetration factor calculated by optical model potential is $0$, if $ \varepsilon _{{m_1}}^c <{V_{Coul}}$.

\subsection{ Double Differential Cross Section of Light Composite Particle}\label{sect2.3}
The double differential cross sections of the emitted neutron and proton can be calculated using the linear momentum-dependent exciton state density model \cite{M.B.Chadwick1991}. The formulas for double differential cross sections of outgoing light composite particles ( deuteron, triton, $^3$He,  $\alpha $ and $^5$He) can be expressed as \cite{Zhang2015}

\begin{eqnarray}\label{eq7}
\frac{{{d^2}\sigma }}{{d\varepsilon d\Omega }} = \sum\limits_n {\frac{{d\sigma (n)}}{{d\varepsilon }}A(n,\varepsilon ,\Omega )} .
\end{eqnarray}
Where $\frac{{d\sigma (n)}}{d\varepsilon }$  is energy spectrum in $n$-th exciton state, and can be calculated with angular momentum and parity conservation. A$(n,\varepsilon ,\Omega )$  is angle factor satisfying the normalization condition, expressed as
\begin{eqnarray}\label{eq8}
A(n,\varepsilon ,\Omega ) = \frac{1}{{4\pi }}\sum\limits_l {(2l + 1)\frac{{{G_l}(\varepsilon ,n)}}{{{G_0}(\varepsilon ,n)}}} \frac{{{\tau _l}(n,\varepsilon )}}{{{\tau _0}(n,\varepsilon )}}{P_l}(\cos \theta ).
\end{eqnarray}
Where $\Omega$ is solid angle of outgoing particle. ${\tau _l}(n,\varepsilon )$  is lifetime of the $l$-th  partial wave with outgoing particle energy $\varepsilon$ emitted from $n$-th exciton state, and can be derived from the exciton model with angular momentum and parity conservation. ${G_l}(\varepsilon ,n)$  is geometric factor in $n$-th exciton state with outgoing particle energy $\varepsilon$, expressed as
\begin{eqnarray}\label{eq9}
G_l^b({\varepsilon _b}) = \frac{1}{{{x_b}}}\int\limits_{\max \left\{ {1,{x_b} - {A_b} + 1} \right\}}^{\sqrt {1 + \frac{E}{{{\varepsilon _F}}}} } {{x_1}d{x_1}} \int\limits_{{x_b} - {x_1}}^{{A_b} - 1} {dy{Z_b}(y){P_l}(\cos \Theta )}.
\end{eqnarray}
Where $\varepsilon _b$ is kinetic energy of the outgoing composite particle, and  $\varepsilon _F$ is Fermi energy. $E^*$ is excited energy of compound nucleus. $A_b$ is mass number of outgoing particle $b$. Here $x_1=p_1/p_F$, $p_1$ is momentum of the first nucleon in the outgoing composite particle $b$, and $p_F$ is Fermi momentum. And $x_b=p_b/p_F$, $p_b$ is momentum of the outgoing composite particle. $y=p_y/p_F$, and $p_y$ is total momentum of nucleons except the first nucleon in the outgoing composite particle $b$. ${Z_b}(y)$ is a factor related to emitted composite particle, expressed as
\begin{eqnarray}\label{eq10}
{Z_b}(y) = \left\{ {\begin{array}{*{20}{l}}
{y, ~~ ~~ ~~ ~~ ~~ ~~ ~~ ~~ ~~ ~~ ~~ ~~ ~~ ~~ ~~ ~~ ~~ ~~ ~~ ~~~~~~~~ ~~ ~~~~~~~~~b = \textmd{deuteron}}\\
{y{{(y - 2)}^2}(y + 4), ~~ ~~ ~~ ~~ ~~ ~~ ~~ ~~ ~~~~~~~~ ~~ ~ ~~~~~~~~ ~~b = \textmd{triton}, \textmd{$^3$He}}\\
{{{(y - 3)}^4}({y^3} + 12{y^2} + 27y - 6),~~~~~~~~~~~~~~~~~~~~b = {\alpha} }\\
{{{(y - 4)}^6}({y^4} + 24{y^3} + 156{y^2} + 224y - 144),~~~~~b = \textmd{$^5$He}}.
\end{array}} \right.
\end{eqnarray}
Cosinoidal function $ \cos \Theta$ can be expressed as
  \begin{eqnarray}\label{eq11}    
\cos\Theta  = \frac{{x_b^2 + x_1^2 - {y^2}}}{{2{x_b}{x_1}}}.  
\end{eqnarray}                                     
The formulas mentioned above can be used to calculate the double differential cross sections of the outgoing neutron, proton, deuteron, triton, $^3$He, $\alpha$, and $^5$He in this work. The calculated results are given in Sec IV.

\section{ANALYSIS OF REACTION CHANNELS FOR $p + ^7$Li REACTION}\label{sect3}
For proton induced $^7$Li reaction, reaction channels exist theoretically at incident energy $E_p \leq 20$ MeV in terms of the reaction threshold energy $E_{th}$ as follows:
\begin{eqnarray}\label{eq12}
p+^7\textmd{Li}\rightarrow ^{8}\textmd{Be}^* \rightarrow \left\{
\begin{array}{llr}
(p, \gamma)^{8}\textmd{Be}, ~~~~~Q=+17.255 \textmd{MeV}, ~~~~~E_{th}=0.000 \textmd{MeV}\\
(p, n)^{7}\textmd{Be},~~~~~Q=-1.643 \textmd{MeV}, ~~~~~~E_{th}=1.879 \textmd{MeV}\\
(p, p)^{7}\textmd{Li},~~~~~~~Q=~0.000 \textmd{MeV}, ~~~~~~E_{th}=0.000 \textmd{MeV}\\
(p, \alpha)\alpha, ~~~~~~~~Q=+17.348 \textmd{MeV}, ~~~~E_{th}=0.000 \textmd{MeV}\\
(p, ^3\textmd{He})^{5}\textmd{He}, ~~~Q=-4.125 \textmd{MeV}, ~~~~~E_{th}=4.7175 \textmd{MeV}\\
(p, d)^{6}\textmd{Li}, ~~~~~~~Q=-5.025 \textmd{MeV}, ~~~~~E_{th}=5.7468 \textmd{MeV}\\
(p, t)^{5}\textmd{Li}, ~~~~~~~Q=-4.434 \textmd{MeV}, ~~~~~E_{th}=5.0709 \textmd{MeV}\\
(p, 2n)^{6}\textmd{Be}, ~~~~Q=-12.320 \textmd{MeV}, ~~~~E_{th}=14.0897 \textmd{MeV}\\
(p, np)^{6}\textmd{Li}, ~~~~~~Q=-7.249 \textmd{MeV}, ~~~~E_{th}=8.2903 \textmd{MeV}\\
(p, pn)^{6}\textmd{Li}, ~~~~~~Q=-7.249 \textmd{MeV}, ~~~~E_{th}=8.2903 \textmd{MeV}\\
(p, n\alpha)^{3}\textmd{He}, ~~~~~Q=-3.230 \textmd{MeV}, ~~~~E_{th}=3.694 \textmd{MeV}\\
(p, nd)^{5}\textmd{Li}, ~~~~~~Q=-10.691 \textmd{MeV}, ~~~E_{th}=12.2267 \textmd{MeV}\\
(p, 2p)^{6}\textmd{He}, ~~~~~~Q=-9.974 \textmd{MeV}, ~~~~E_{th}=11.4067 \textmd{MeV}\\
(p, pt)^{4}\textmd{He}, ~~~~~~Q=-2.467 \textmd{MeV}, ~~~~E_{th}=2.8214 \textmd{MeV}\\
(p, tp)^{4}\textmd{He}, ~~~~~~Q=-2.467 \textmd{MeV}, ~~~~E_{th}=2.8214 \textmd{MeV}\\
(p, pd)^{5}\textmd{He}, ~~~~~~Q=-9.619 \textmd{MeV}, ~~~E_{th}=11.0007 \textmd{MeV}\\
(p, dp)^{5}\textmd{He}, ~~~~~~Q=-9.619 \textmd{MeV}, ~~~E_{th}=11.0007 \textmd{MeV}.
\end{array}
\right.
\end{eqnarray}
Considering the conservations of the energy, angular momentum, and parity in the particle emission processes, the reaction channels of the first particle emission are listed as follows:
\begin{eqnarray}\label{eq13}
p+^7\textmd{Li}\rightarrow ^{8}\textmd{Be}^* \rightarrow \left\{
\begin{array}{l}
n+ ^{7}\textmd{Be}^* ~~(k_1=gs, 1, 2, ...,7),\\
p+ ^{7}\textmd{Li}^* ~~(k_1=gs, 1, 2, ..., 10),\\
\alpha+{\alpha}^* ~~(k_1=gs, 1, 2, ..., 14),\\
^3\textmd{He}+ ^{5}\textmd{He}^* ~~(k_1=gs, 1),\\
d+ ^6\textmd{Li}^* ~~(k_1=gs, 1, 2, ..., 5),\\
t+ ^5\textmd{Li}^* ~~(k_1=gs,1,2).
\end{array}
\right.
\end{eqnarray}
Where $gs$ and $k_1$ denote the ground state and the $k_1$-th energy level of the first residual nuclei $M_1$ taken from measurements \cite{Tilley1992,Tilley2002,Tilley2004}, respectively.

For the first particle emission channel $^7$Li($p, n$)$^7$Be$^*$, the first residual nucleus $^7$Be$^*$, which attains the seventh energy level, can still emit a proton with the residual nucleus $^6$Li or a alpha particle with the residual nucleus $^3$He. Furthermore, the secondary residual nucleus $^6$Li can break up into deuteron and alpha particle \cite{Zhang2001Li6} if $^6$Li is in the first, third and fourth discrete energy levels. Therefore, the first particle emission channel $^7$Li($p, n$)$^7$Be$^*$ can further open ($p, np$)$^6$Li, ($p, npd\alpha$) and ($p, n\alpha$)$^3$He reaction channels in the final state.

For the first particle emission channel $^7$Li($p, p$)$^7$Li$^*$, the first excited level of residual nucleus $^7$Li cannot emit any particle, so it purely contributes to the inelastic scattering channel. The second and the third excited levels of $^7$Li can emit triton, so they contribute to the ($p,  pt\alpha$) reaction channel. If the first residual nucleus $^7$Li$^*$ is at the $k_1$-th ($k_1 \geq 4$) excited energy level, some
energy levels will emit a neutron, so they contribute to the $(p,  pn)^6$Li reaction channel. Furthermore, the secondary residual nucleus $^6$Li with high excited energy can break up into $d + \alpha$, thus this reaction process belongs to ($p, pnd\alpha$) reaction channel. If the first residual nucleus $^7$Li$^*$ is at the $k_1$-th ($k_1 \geq 6$) excited energy level, some energy levels will emit proton and deuteron with the corresponding secondary residual nuclei as $^6$He$_{gs}$ and $^5$He, respectively. Considering the two cluster separation reaction, i.e. $^5$He $\rightarrow$ $n + \alpha$ \cite{Yan 5He}, so these reaction processes belong to ($p,2p$)$^6$He$_{gs}$ and ($p,pnd\alpha$) reaction channels, respectively. Therefore, first particle emission channel $^7$Li($p, p$)$^7$Li$^*$ will contribute to ($p, pn$)$^6$Li, ($p, pt\alpha$), ($p, npd\alpha$) and ($p, 2p$)$^6$He$_{gs}$ reaction channels in the final state, besides the elastic and inelastic scattering.

For the first particle emission channel $^7$Li($p,d$)$^6$Li$^*$, besides reaction process $^6$Li$^*$ $\rightarrow$ $d + \alpha$ as mentioned above belongs to ($p, 2d\alpha$) reaction channel in the final state, some excited energy levels ($k_1 > 3$) of the first residual nucleus $^6$Li$^*$ can emit proton with the secondary residual nucleus $^5$He. As mentioned above, $^5$He is unstable and can be separated into a neutron and an alpha particle spontaneously \cite{Yan 5He}, so ($p, dp$)$^5$He reaction channel belongs to ($p, pnd\alpha$) reaction channel in the final state.  

Considering the two-cluster separation processes, i.e. $^5$Li $\rightarrow$ $p + \alpha$ and $^5$He$ \rightarrow$ $n + \alpha$, so the first particle emission channels such as ($p, t$)$^5$Li and ($p, $$^3$He)$^5$He belong to ($p, pt\alpha$) and ($p, n\alpha$)$^3$He reaction channels in the final state, respectively.

For proton induced $^7$Li reaction at $E_p=14$ MeV, the compound nucleus $^8$Be can even reach the twenty-seventh discrete energy level with $28.6$ MeV in term of Eq. (\ref{eq4}), so it can emit neutron, proton, deuteron, triton, $^3$He, and break up into two alpha particles. Because of the high excited energy of the compound nucleus $^8$Be, alpha particles through two body breakup process are also at high excited energy in term of the energy conservation. So $^4$He can emit a proton at $k_1$-th ($k_1\geq 1$) energy level, a neutron at $k_1$-th ($k_1\geq 2$) energy level, and break up into two deuterons at $k_1$-th ($k_1\geq 9$) energy level, respectively. These reaction processes belong to ($p, 2\alpha$), ($p, pt\alpha$), ($p,n\alpha$)$^3$He and ($p, 2d\alpha$) reaction channels. Certainly, the gamma decay obviously competes with the particle emission, and the branching ratios can be obtained by means of the model calculation in STLN. 

According to the analysis of reaction channels mentioned above, the total spectra could be produced by adding all of the partial spectra of the same outgoing particle yielded from every reaction channel. The contributions of the double differential cross sections of total emitted proton are from elastic scattering, inelastic scattering, ($p,np$)$^6$Li, ($p,2p$)$^6$He$_{gs}$, ($p, npd\alpha$) and ($p, pt\alpha$) reaction channels. The contributions of the double differential cross sections of total emitted deuteron are from ($p,d$)$^6$Li, ($p, npd\alpha$) and ($p, 2d\alpha$) reaction channels. The contributions of the double differential cross sections of total emitted triton are just from $(p,t)^5$Li and ($p, pt\alpha$) reaction channels. The contributions of the double differential cross sections of total emitted neutron are just from ($p,n$)$^7$Be, ($p, n^3$He)$\alpha$, $(p,npd\alpha)$ and $(p,np)^6$Li reaction channels. The contribution of the double differential cross sections of total emitted $^3$He is only from ($p, n^3$He)$\alpha$ reaction channel. In conclusion, for the proton induced $^7$Li reaction, reaction channels
exist at incident energy $E_p \leq 20$ MeV as follows:
\begin{eqnarray}\label{eq14}
p + {}^7\textmd{Li} \to {}^8\textmd{Be}^* \to \left\{ {\begin{array}{*{20}{l}}
{n + {}^7\textmd{Be}^*\left\{ {\begin{array}{*{20}{l}}
{k_1 = gs,1~~~~~~~~~~~~~~~~~~~~~~~~~(p,n){}^7\textmd{Be}}\\
{k_1 = 2,7~~~~~~~~~~~~~~~~~~~~~~~~~(p,n{}^3\textmd{He})\alpha }\\
{k_1 = 4,7~~~~~~~~~~~~~~~~~~~~~~~~(p,np){}^6\textmd{Li}_{gs}}
\end{array}} \right.}\\
{p + {}^7\textmd{Li}^*\left\{ {\begin{array}{*{20}{l}}
{k_1 = gs ~~~~~~~~~~~~~~~~Compound~elastic}\\
{k_1 = 1 ~~~~~~~~~~~~~~~~~~~~~~~~~~~~~~~~~~(p,{p'})}\\
{k_1 = 2,...,10~~(t + \alpha )~~~~~~~~~~~~~~(p,pt\alpha )}\\
{k_1 \ge 6(d + {}^5\textmd{He})~~~~~~~~~~~~~~~~~~(p,npd\alpha )}\\
{k_1 \ge 6(p + {}^6\textmd{He}_{gs})~~~~~~~~~~~~(p,2p){}^6\textmd{He}_{gs}}\\
{k_1 \ge 4(n + {}^6\textmd{Li})~~~~~~~~~~~~~~~~~(p,np){}^6\textmd{Li}}
\end{array}} \right.}\\
{\alpha  + {\alpha ^*}\left\{ {\begin{array}{*{20}{l}}
{k_1 = gs~~~~~~~~~~~~~~~~~~~~~~~~~~~~~~~~~(p,2\alpha )}\\
{k_1 = 1 ~~~~~~~~~~~~~~~~~~~~~~~~~~~~~~~~~~(p,pt\alpha )}\\
{k_1 \ge 2~~~~~~~~~~~~~~~~~~~(p,pt\alpha ),(p,n{}^3\textmd{He}\alpha )}\\
{k_1 \ge 9~~~~~~~~(p,pt\alpha ),(p,n{}^3\textmd{He}\alpha ),(p,2d\alpha )}
\end{array}} \right.}\\
{{}^3\textmd{He} + {}^5\textmd{He}~(k_1 = gs,1) \to (n + \alpha )~~~~~~~~(p,n{}^3\textmd{He}\alpha )}\\
{d + {}^6\textmd{Li}^*\left\{ {\begin{array}{*{20}{l}}
{k_1 = gs,2~~~~~~~~~~~~~~~~~~~~~~~~~~~~~~~~(p,d)}\\
{k_1 = 1,3,4,5~~~~~~~~~~~~~~~~~~~~~~~~~(p,2d\alpha )}\\
{k_1 = 4,5(p+^5\textmd{He} \to n + \alpha)~~~~(p,npd\alpha)}
\end{array}} \right.}\\
{t + {}^5\textmd{Li}~(k_1 = gs,1,2) \to (p + \alpha )~~~~~~~~~~~~~~(p,pt\alpha )}.
\end{array}} \right.
\end{eqnarray}

\section{THE CALCULATED RESULTS AND ANALYSIS}\label{sect4}
The experimental double differential cross sections of proton for $p + ^7$Li reaction had been measured only at incident proton energy $E_p = 14$ MeV in 1989 \cite{N. Koori1989}. The experimental double differential cross sections of deuteron and triton for $p + ^7$Li reaction had also been given in 1991 \cite{N. Koori1991}. The PUNF code based on STLN is developed for calculating the cross sections, elastic angular distributions and the double differential cross sections of outgoing neutrons, proton and light charged particles. In this paper, the comparisons between the calculated results with the measurements of double differential cross sections of total outgoing proton, deuteron and triton for $p + ^7$Li reaction will be performed. 

\begin{figure}
\centering
\includegraphics[width=15cm,angle=0]{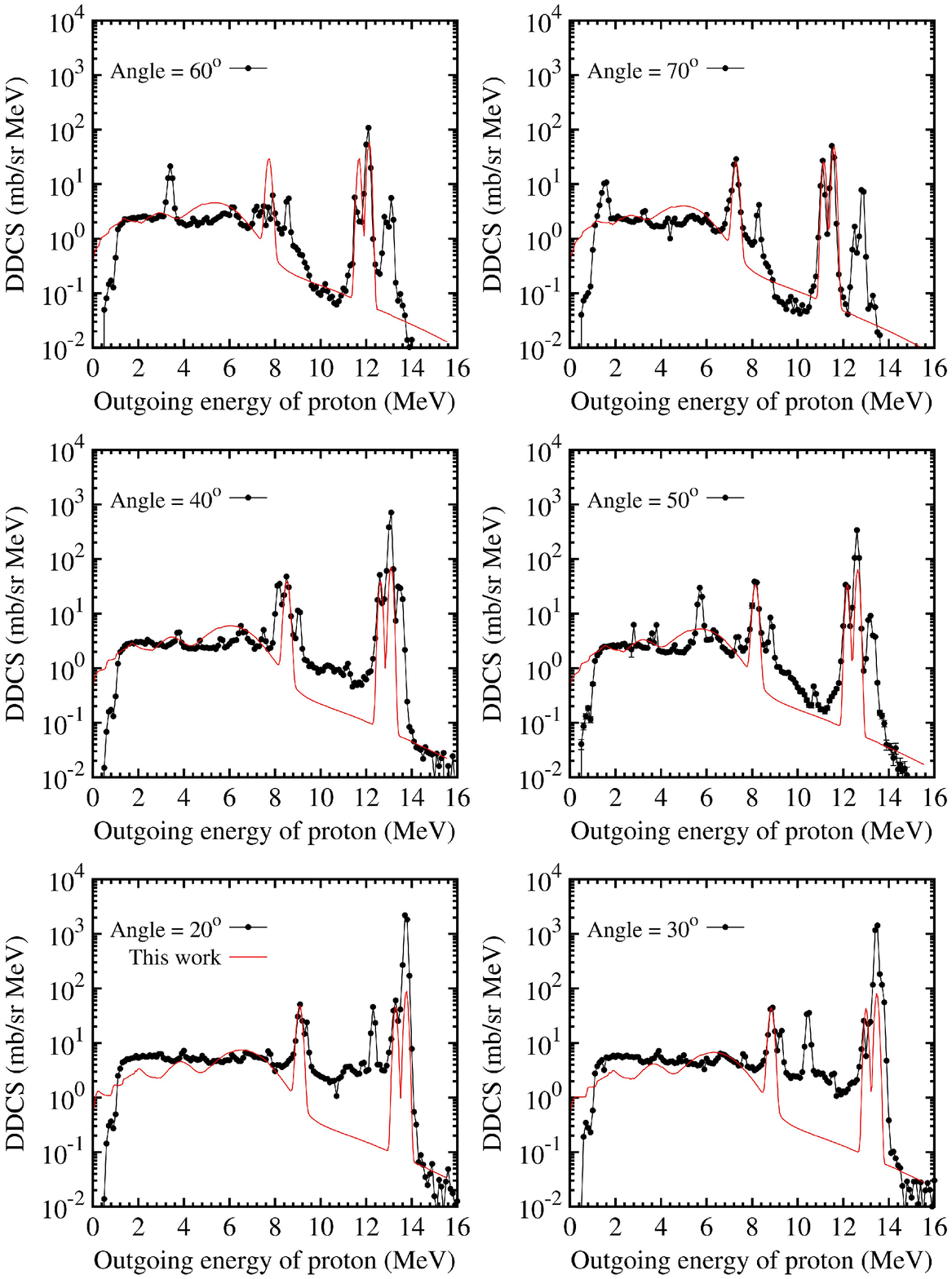}
\caption{(Color online)  The total double differential cross sections of outgoing proton for $p + ^7$Li reaction with outgoing angles $20^\circ, 30^\circ,40^\circ, 50^\circ, 60^\circ$ and $70^\circ$ at $E_p = 14$ MeV in LS. The black points denote the experimental data taken from Ref. \cite{N. Koori1989}. The red solid lines denote the calculated results.  }\label{Fig1}
\end{figure}

\begin{figure}
\centering
\includegraphics[width=15cm,angle=0]{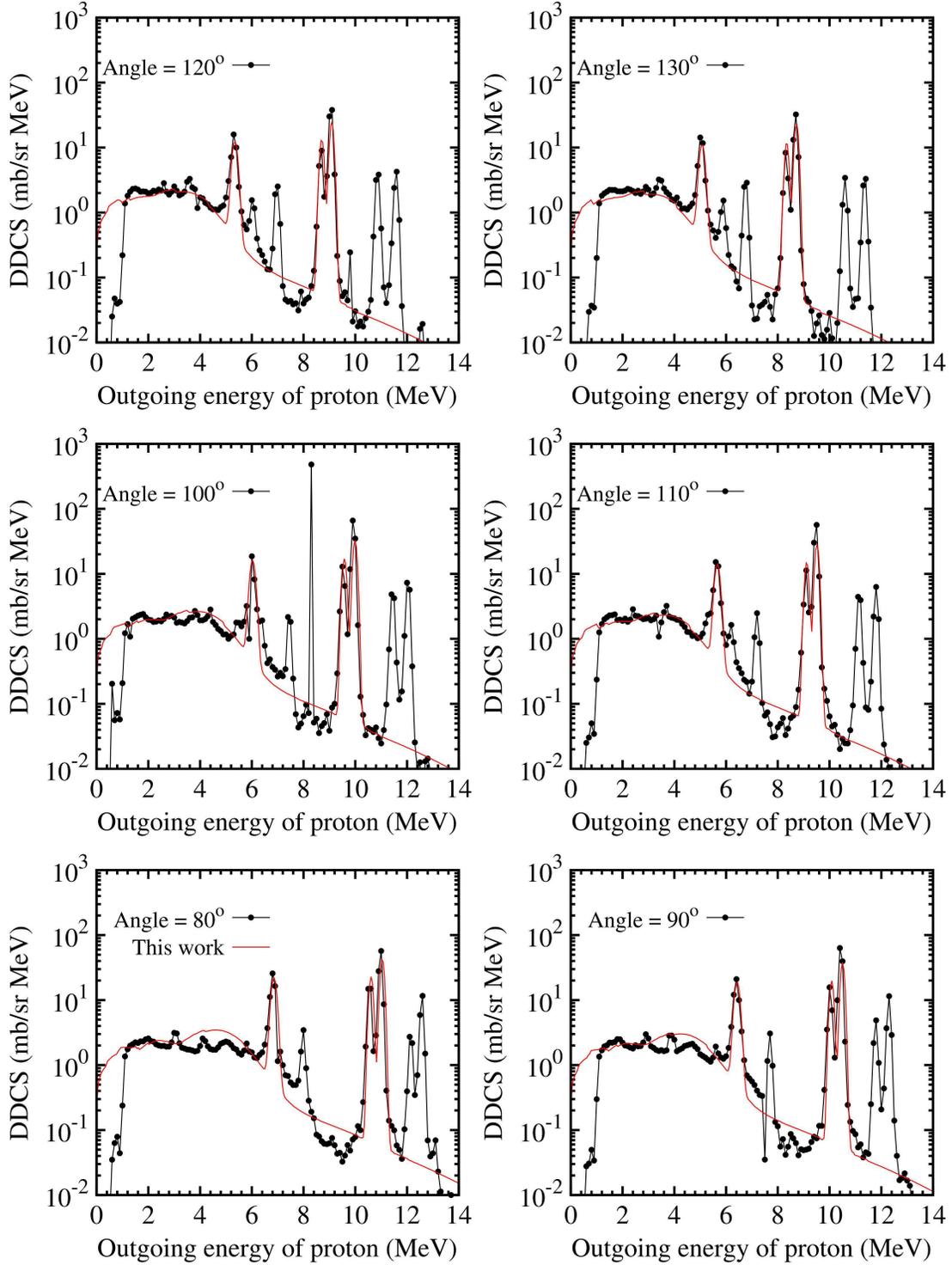}
\caption{(Color online)The same as Fig. \ref{Fig1}, but at outgoing angles $80^\circ, 90^\circ,100^\circ, 110^\circ, 120^\circ$ and $130^\circ$. }\label{Fig2}
\end{figure}

\begin{figure}
\centering
\includegraphics[width=15cm,angle=0]{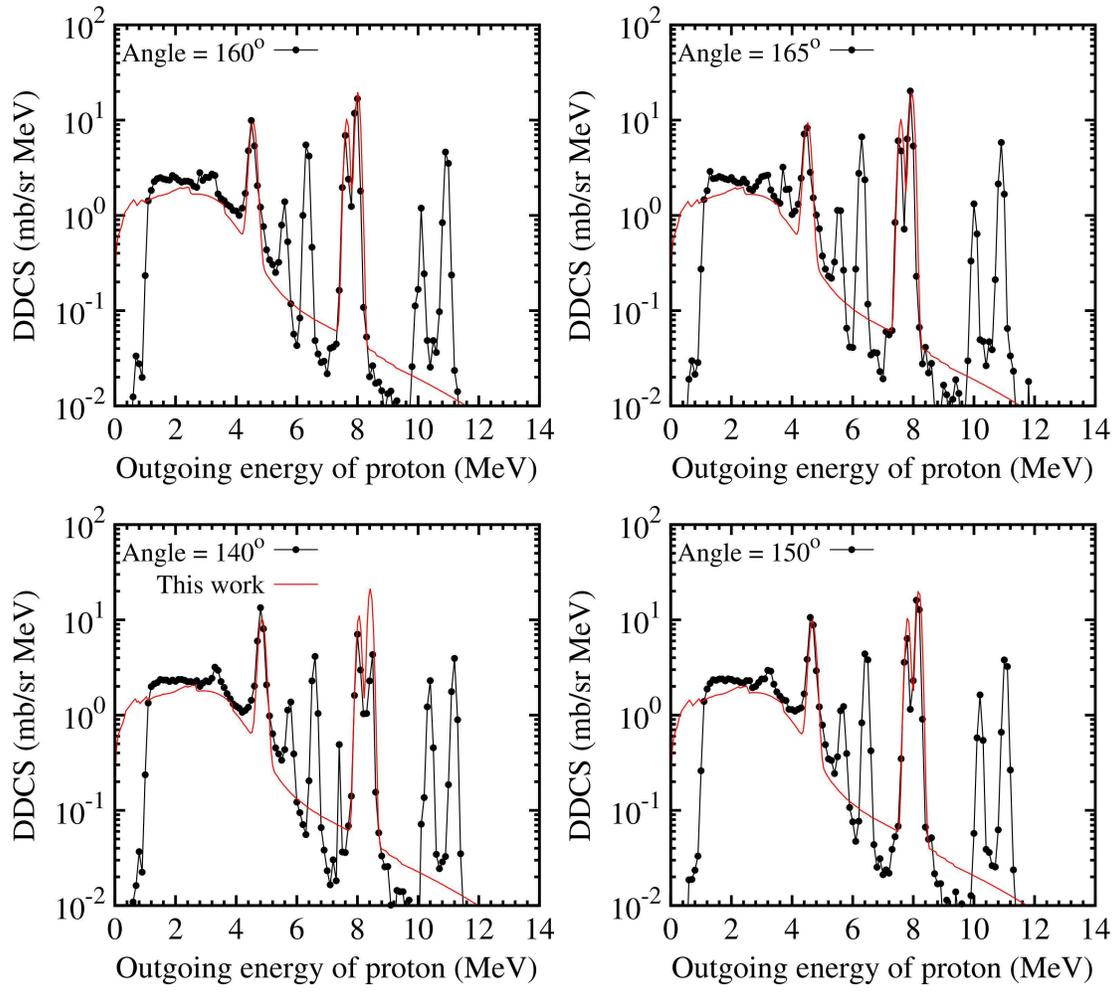}
\caption{(Color online)The same as Fig. \ref{Fig1}, but at outgoing angles $140^\circ, 150^\circ,160^\circ$ and $165^\circ$.}\label{Fig3}
\end{figure}

\begin{figure}
\centering
\includegraphics[width=20cm,angle=0]{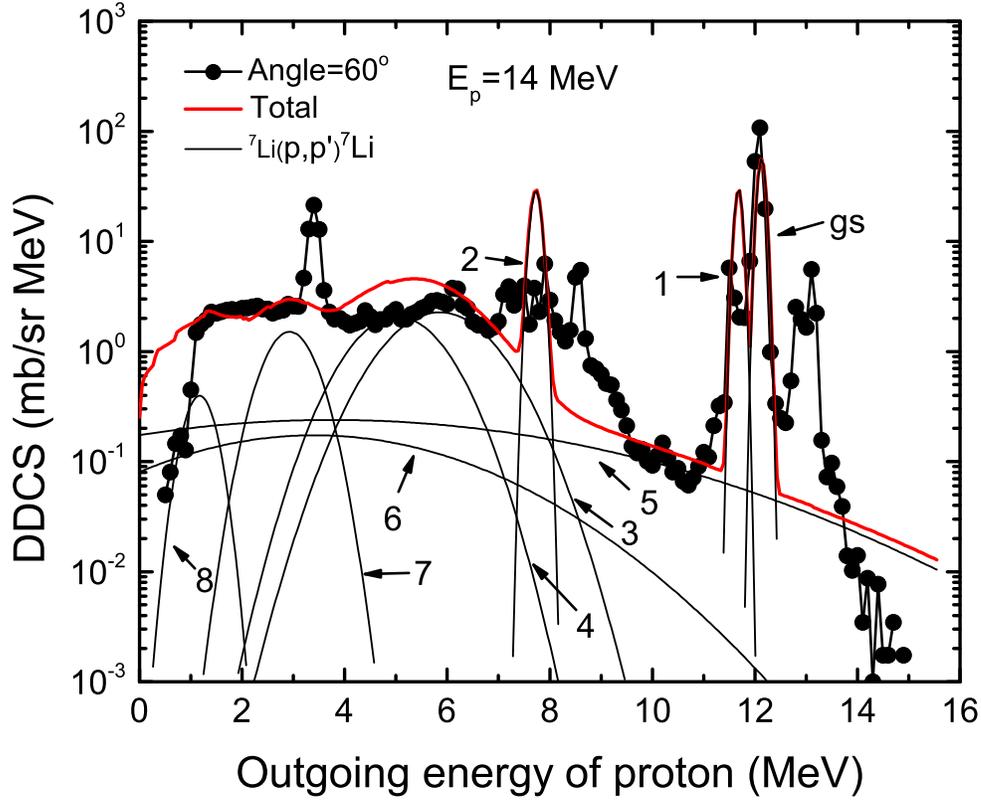}
\caption{\label{Fig_Par_P1}(Color online) The partial double-differential cross sections of outgoing proton from reaction channel ($p,{p'}$)$^7$Li with outgoing angle $60^\circ$ at $E_p = 14$ MeV in LS. The red solid line denotes the calculated total double-differential cross sections. The black solid lines denote the partial spectra of the first outgoing proton from the compound nucleus to the ground state, up to the eighth excited energy levels (as labeled in the figure) of the first residual nucleus $^7$Li. Only the cross sections with values larger than 10$^{-3}$ mb are given.}
\end{figure}

\begin{figure}
\centering
\includegraphics[width=20cm,angle=0]{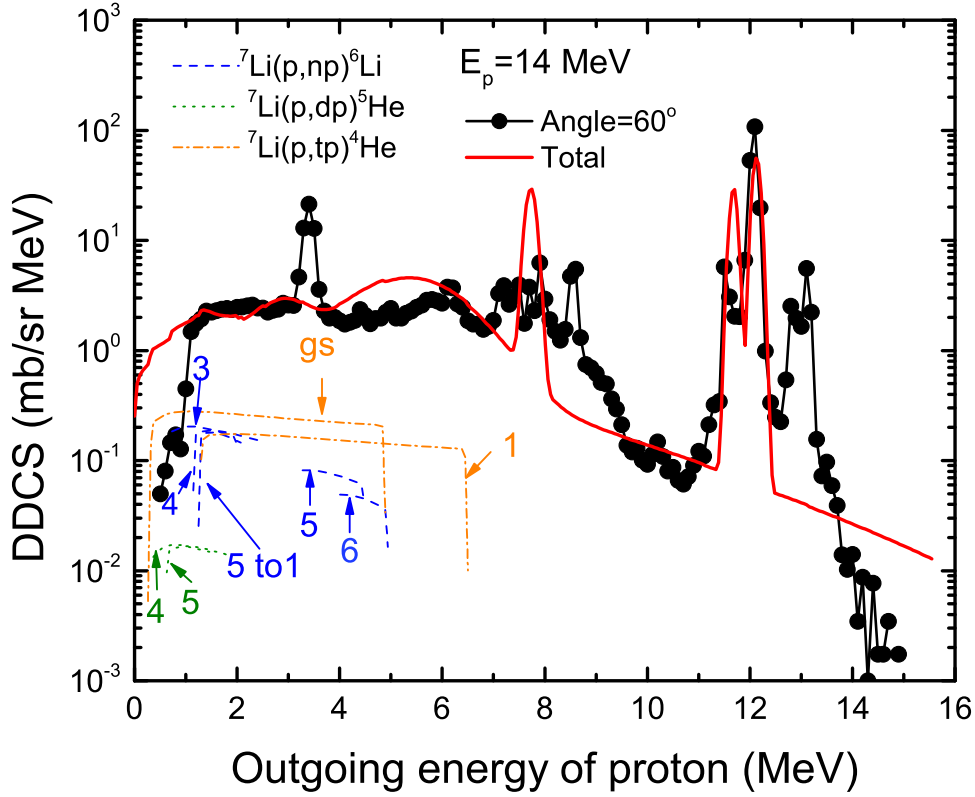}
\caption{\label{Fig_Par_P2}(Color online) The same as Fig. \ref{Fig_Par_P1}, but the blue dashed lines denote the partial spectra of secondary outgoing proton from the third to sixth excited energy levels of $^7$Be to the ground state of $^6$Li, and the fifth excited energy level of $^7$Be to the first excited energy level of $^6$Li for $^7$Li($p, np$)$^6$Li reaction channel. The green dotted lines denote the partial spectra of secondary outgoing proton from the fourth and fifth excited energy levels of $^6$Li to the ground state of $^5$He for $^7$Li($p, dp$)$^5$He reaction channel. The orange dash-dotted lines denote the partial spectra of secondary outgoing proton from ground state and the first excited energy level of $^5$Li to ground state of $^4$He for $^7$Li($p, tp$)$^4$He reaction channel.}
\end{figure}

\begin{figure}
\centering
\includegraphics[width=20cm,angle=0]{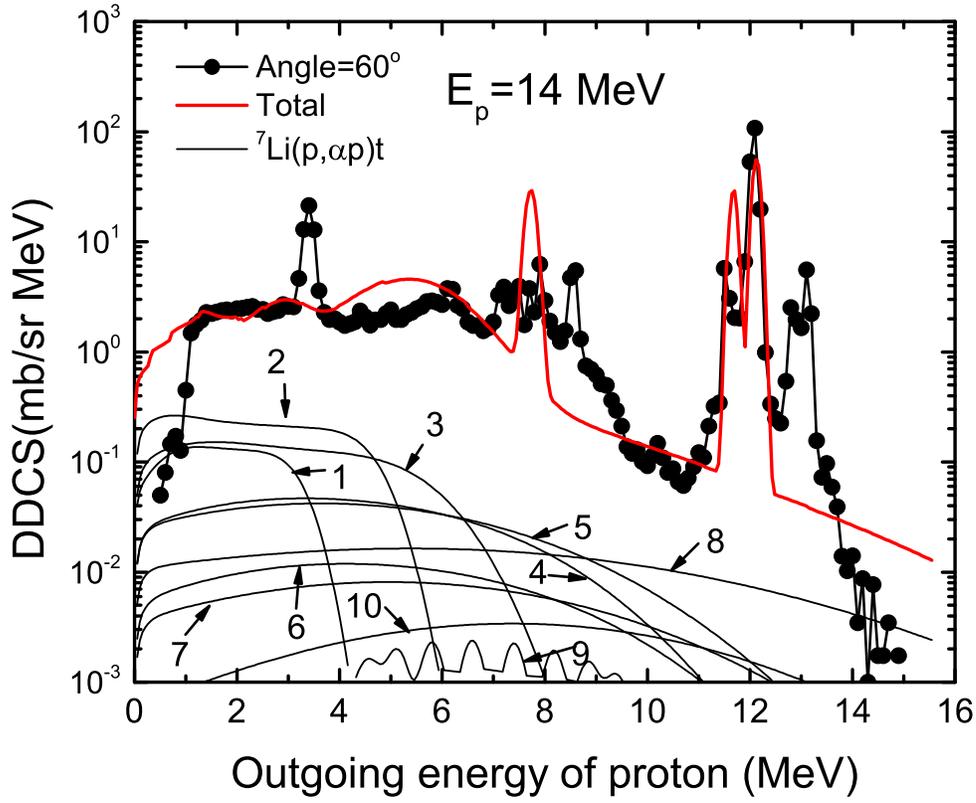}
\caption{\label{Fig_Par_P3}(Color online)  The same as Fig. \ref{Fig_Par_P1}, but the black solid lines denote the partial spectra of secondary outgoing proton from the first to $10$th excited energy levels of $^4$He, which can break into $p$ and $t$ for $^7$Li($p, \alpha p)t$ reaction channel. }
\end{figure}

The comparisons of the calculated double differential cross sections of total outgoing proton with the measured data are shown in Figs. \ref{Fig1} - \ref{Fig3} at the incident proton energy $14$ MeV for outgoing angles $20^\circ, 30^\circ, 40^\circ, 50^\circ, 60^\circ, 70^\circ, 80^\circ, 90^\circ, 100^\circ, 110^\circ, 120^\circ, 130^\circ, 140^\circ, 150^\circ, 160^\circ $ and $165^\circ$, respectively. The black points denote the experimental data derived from Ref. \cite{N. Koori1989}, and the red solid lines denote the calculated total double differential cross sections of outgoing proton. The calculated results agree well with the measurements except some peaks which are contaminated by the scattering from $^1$H, $^{12}$C and $^{16}$O as illuminated in Ref. \cite{H.R.Guo2019,N. Koori1989}.  Fig. \ref{Fig_Par_P1} shows the partial double differential cross sections of outgoing proton from reaction channel $^7$Li($p,{p'}$)$^7$Li with outgoing angle $60^\circ$ at $E_p = 14$ MeV in LS. The black solid lines denote the partial spectra of the first outgoing proton from the compound nucleus to the ground state, up to the eighth excited energy levels of the first residual nucleus $^7$Li as labeled in Fig. \ref{Fig_Par_P1}. In this paper, only the cross sections with values larger than 10$^{-3}$ mb are given. Fig. \ref{Fig_Par_P2} shows the partial spectra of secondary outgoing proton from the third to sixth excited energy levels of $^7$Be to the ground state of $^6$Li, and from the fifth excited energy level of $^7$Be to the first excited energy level of $^6$Li for $^7$Li($p, np$)$^6$Li reaction channel labeled by the blue dashed lines. The green dotted lines denote the partial spectra of secondary outgoing proton from the fourth and fifth excited energy levels of $^6$Li to the ground state of $^5$He for $^7$Li($p, dp$)$^5$He reaction channel. The orange dash-dotted lines denote the partial spectra of secondary outgoing proton from ground state and the first excited energy level of $^5$Li to ground state of $^4$He for $^7$Li($p, tp$)$^4$He reaction channel. Fig. \ref{Fig_Par_P3} shows the partial spectra of secondary outgoing proton from the first to 10th excited energy levels of $^4$He which breaks into $p$ and $t$ for $^7$Li($p, \alpha p)t$ reaction channel labeled by the black solid lines. From Figs. \ref{Fig_Par_P1} - \ref{Fig_Par_P3}, one can see that the contributions of the secondary outgoing proton is far smaller than that of the first outgoing proton, as shown in Ref. \cite{H.R.Guo2019}.

\begin{figure}
\centering
\includegraphics[width=15cm,angle=0]{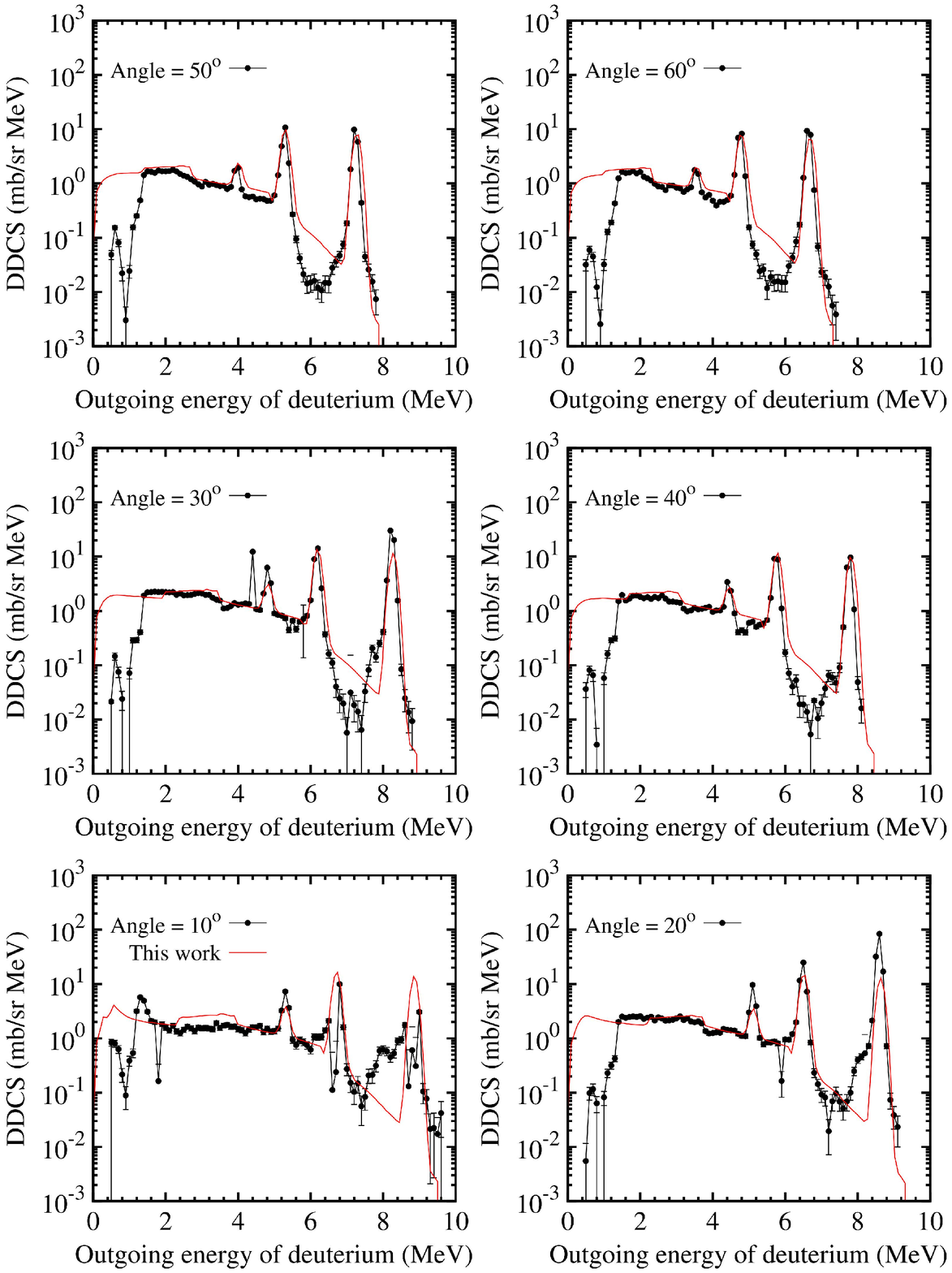}
\caption{(Color online) The same as Fig. \ref{Fig1}, but for outgoing deuteron. The experimental data are derived from Ref. \cite{N. Koori1991}.}\label{Fig4}
\end{figure}

\begin{figure}
\centering
\includegraphics[width=15cm,angle=0]{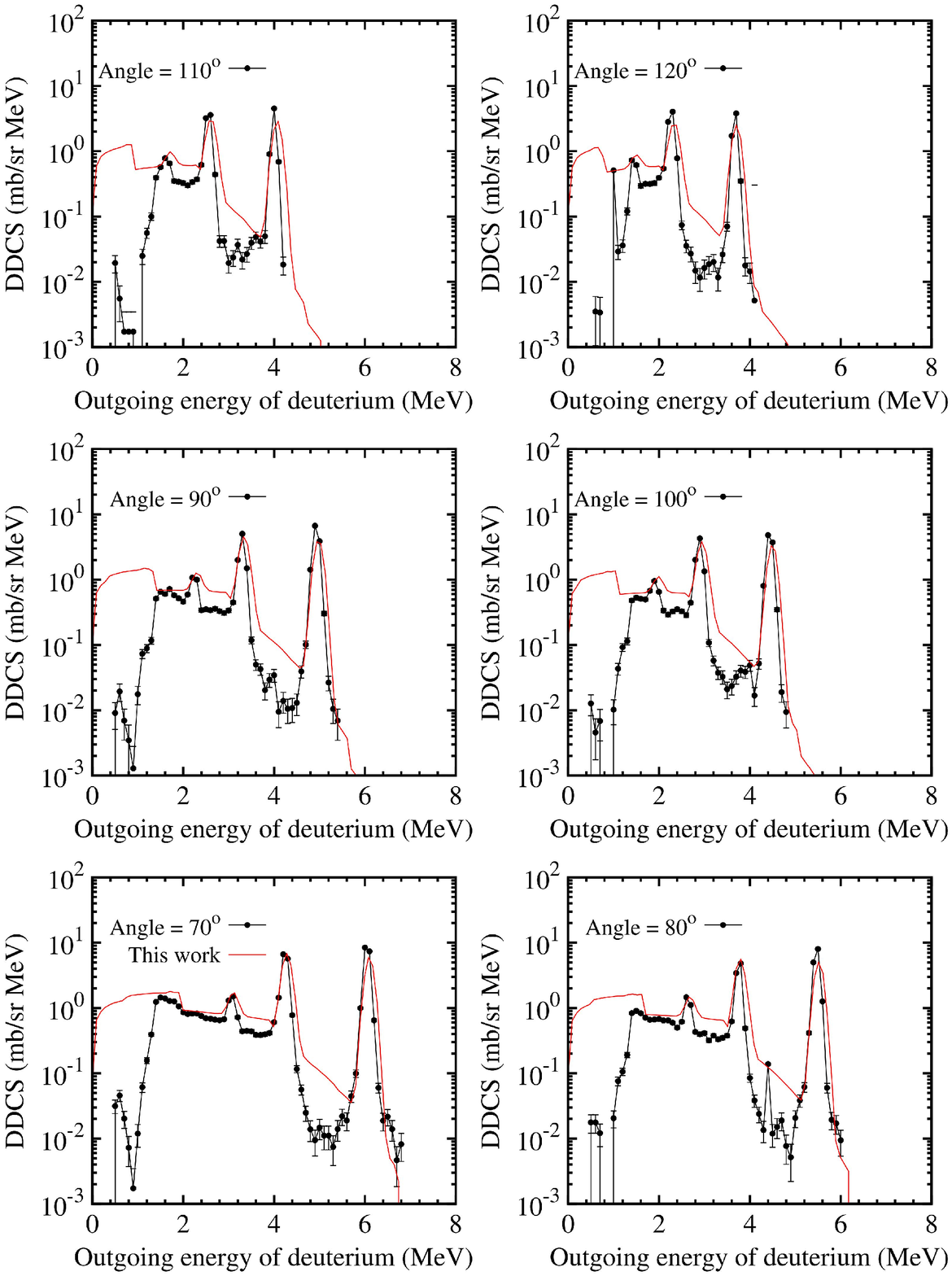}
\caption{(Color online) The same of Fig. \ref{Fig4}, but for different outgoing angles as labeld in this figure.}\label{Fig5}
\end{figure}

\begin{figure}
\centering
\includegraphics[width=15cm,angle=0]{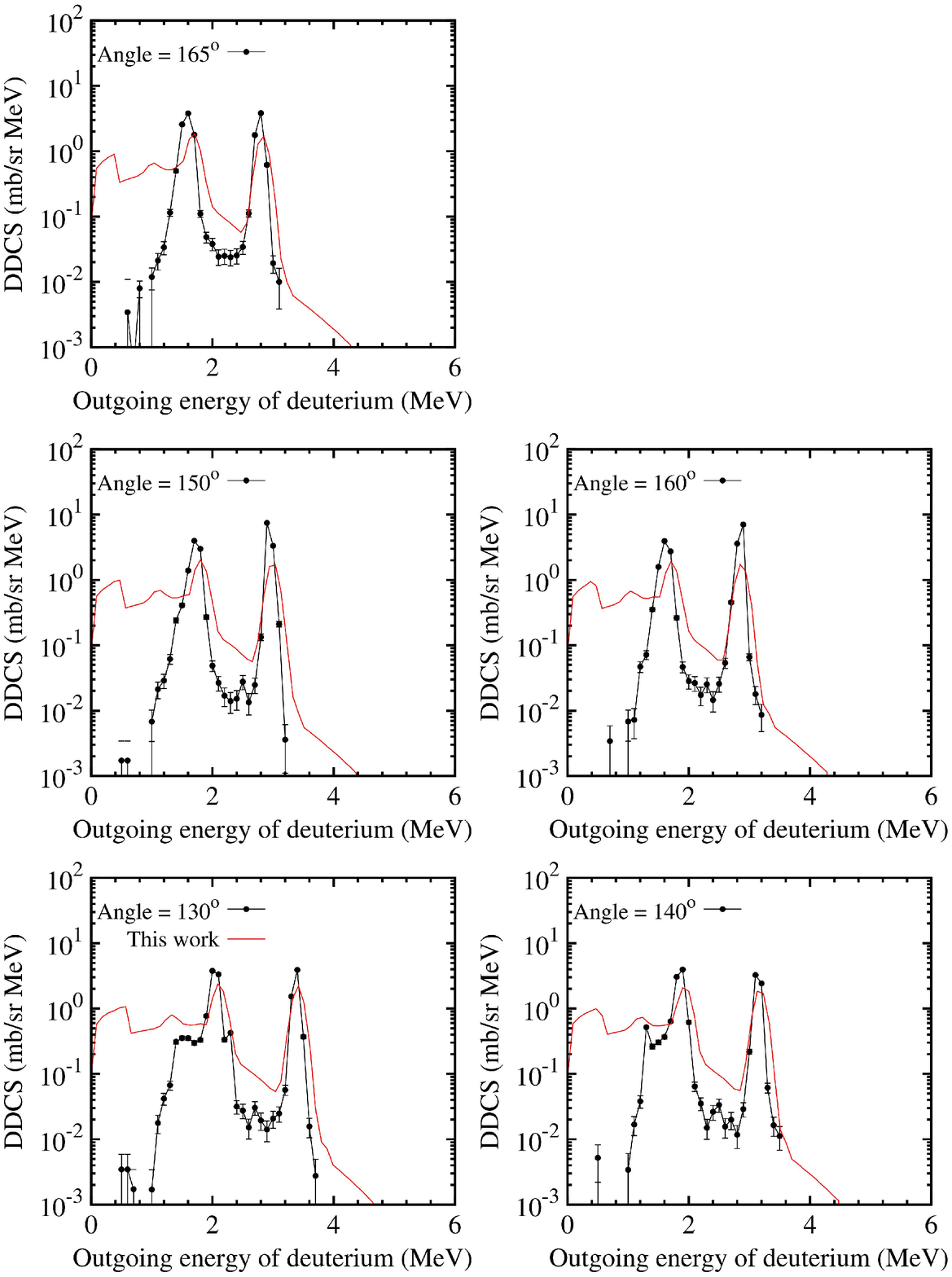}
\caption{(Color online) The same of Fig. \ref{Fig4}, but for different outgoing angles as labeld in this figure.}\label{Fig6}
\end{figure}

\begin{figure}
\centering
\includegraphics[width=20cm,angle=0]{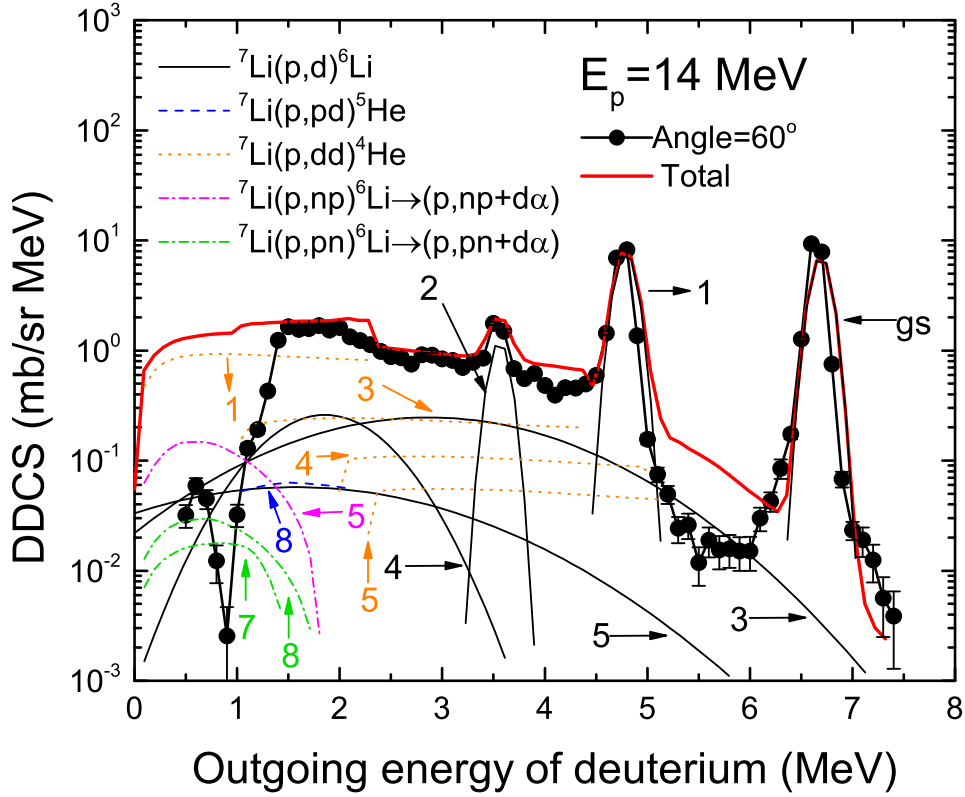}
\caption{\label{Fig_Par_D1}(Color online)  The same as Fig. \ref{Fig_Par_P1}, but for outgoing deuteron. The black solid lines denote the partial spectra of first outgoing deuteron from compound nucleus to the ground state, up to the fifth excited energy level of the first residual nucleus $^6$Li for $^7$Li$(p, d)^6$Li reaction channel. The blue dashed line denotes the contribution from the eighth excited energy level of $^7$Li to the ground state of $^5$He for $^7$Li$(p, pd)^5$He reaction channel. The orange dotted lines denote the contributions from the first, third, fourth and fifth excited energy levels of $^6$Li to the ground state of $^4$He for $^7$Li$(p, dd)^4$He reaction channel. The magenta dash-dotted line denotes the contribution of reaction channel ($p, np$)$^6$Li $\rightarrow$ ($p, np+d\alpha$) from the fifth excited energy level of $^7$Be to the first excited energy level of $^6$Li, which can break up into $d + \alpha$. The green dash-dotted lines denote the contributions of reaction channel ($p, pn$)$^6$Li $\rightarrow$ ($p, pn+d\alpha$) from the seventh and eighth excited energy levels of $^7$Li to the first excited energy level of $^6$Li. }
\end{figure}

\begin{figure}
\centering
\includegraphics[width=20cm,angle=0]{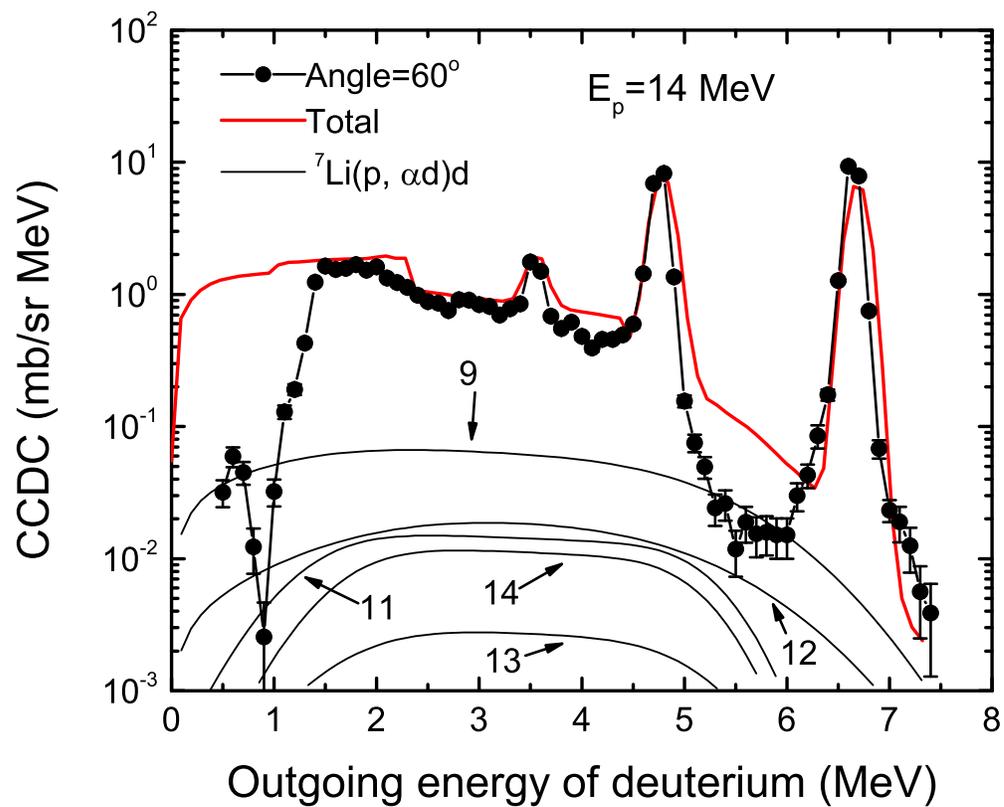}
\caption{\label{Fig_Par_D2}(Color online)  The same as Fig. \ref{Fig_Par_D1}, but for the secondary outgoing deuteron from the 9th to $13$th excited energy levels of $^4$He, which can further break up into 2$d$.  }
\end{figure}

The calculated double differential cross sections of total outgoing deuteron for $p + ^7$Li reaction at $14$ MeV are compared with the experimental data with outgoing angles of $10^\circ, 20^\circ, 30^\circ, 40^\circ, 50^\circ, 60^\circ, 70^\circ, 80^\circ, 90^\circ, 100^\circ, 110^\circ, 120^\circ, 130^\circ, 140^\circ, 150^\circ, 160^\circ$ and $165^\circ$, as shown in Figs. \ref{Fig4} - \ref{Fig6}. The black points denote the experimental data derived from Ref. \cite{N. Koori1991}, and the red solid lines denote the calculated total double-differential cross sections of outgoing deuteron. One can see that the calculated results agree well with the measurements. Fig. \ref{Fig_Par_D1} shows the partial double differential cross sections of outgoing deuteron from reaction channels $^7$Li($p,d$)$^6$Li, $^7$Li($p,pd$)$^5$He, $^7$Li($p,dd$)$^4$He and $^7$Li($p,npd\alpha$) with outgoing angle $60^\circ$ at $E_p = 14$ MeV in LS. In Fig. \ref{Fig_Par_D1}, the black solid lines denote the partial spectra of first outgoing deuteron from compound nucleus to the ground state, up to the fifth excited energy level of the first residual nucleus $^6$Li for $^7$Li$(p, d)^6$Li reaction channel. The blue dashed line denotes the contribution of secondary outgoing deteron from the eighth excited energy level of $^7$Li to the ground state of $^5$He for $^7$Li$(p, pd)^5$He reaction channel. The orange dotted lines denote the contributions of secondary outgoing deteron from the first, third, fourth and fifth excited energy levels of $^6$Li to the ground state of $^4$He for $^7$Li$(p, dd)^4$He reaction channel. The magenta dash-dotted line denotes the contribution of secondary outgoing deteron for reaction channel ($p, np$)$^6$Li $\rightarrow$ ($p, np+d\alpha$) from the fifth excited energy level of $^7$Be to the first excited energy level of $^6$Li, which can break up into $d + \alpha$. The green dash-dotted lines denote the contributions of secondary outgoing deteron for reaction channel ($p, pn$)$^6$Li $\rightarrow$ ($p, pn+d\alpha$) from the seventh and eighth excited energy levels of $^7$Li to the first excited energy level of $^6$Li. Fig. \ref{Fig_Par_D2} shows the partial double differential cross sections of outgoing deuteron from reaction channel $^7$Li($p,\alpha d$)$d$ with outgoing angle $60^\circ$ at $E_p = 14$ MeV in LS. In Fig. \ref{Fig_Par_D2}, the black solid lines denote the secondary outgoing deuteron from the ninth to $13$th excited energy levels of $^4$He, which can further break up into 2$d$. 

\begin{figure}
\centering
\includegraphics[width=15cm,angle=0]{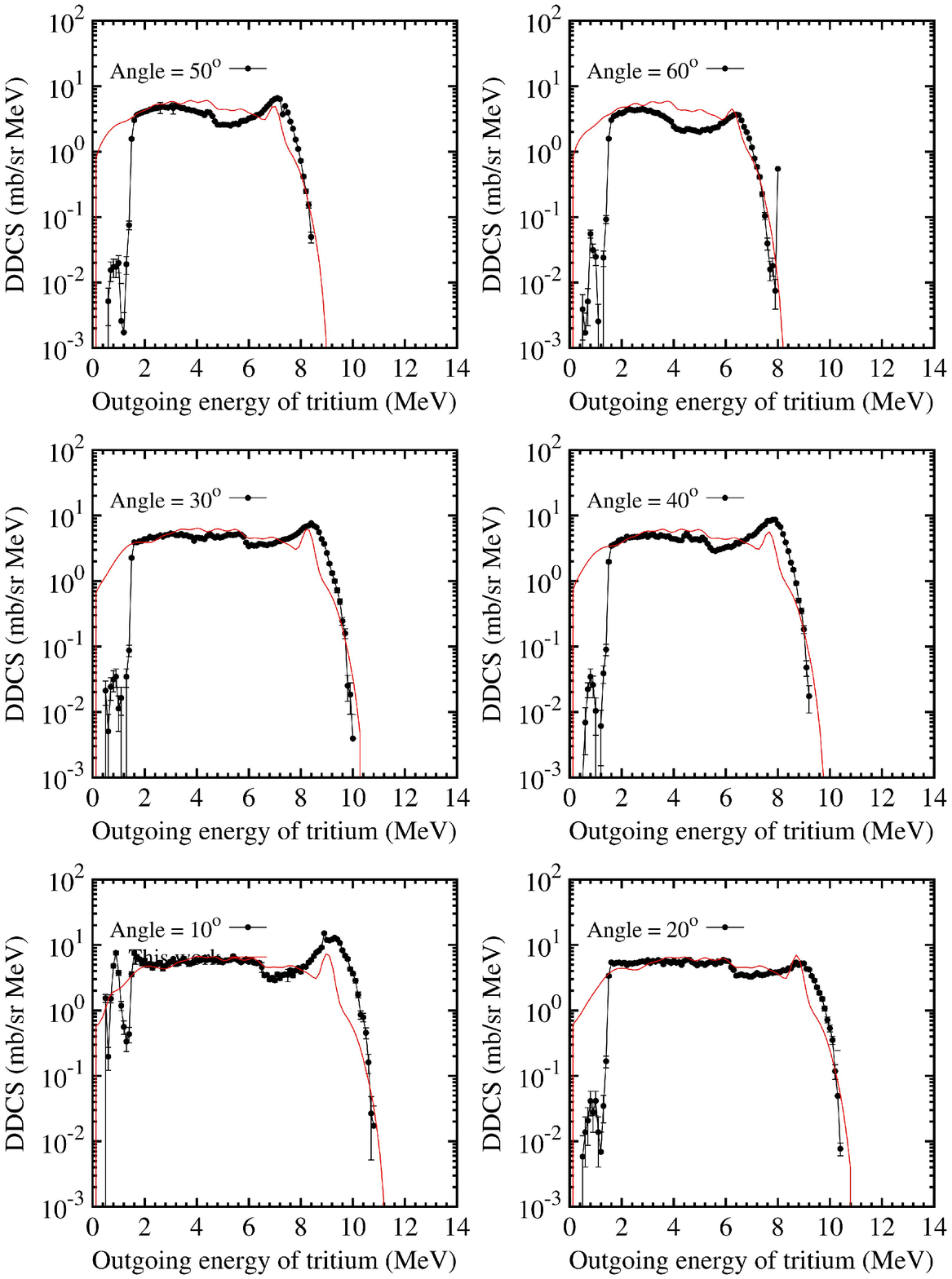}
\caption{(Color online)  The same as Fig. \ref{Fig1}, but for outgoing triton. The experimental data are derived from Ref. \cite{N. Koori1991}.}\label{Fig7}
\end{figure}

\begin{figure}
\centering
\includegraphics[width=15cm,angle=0]{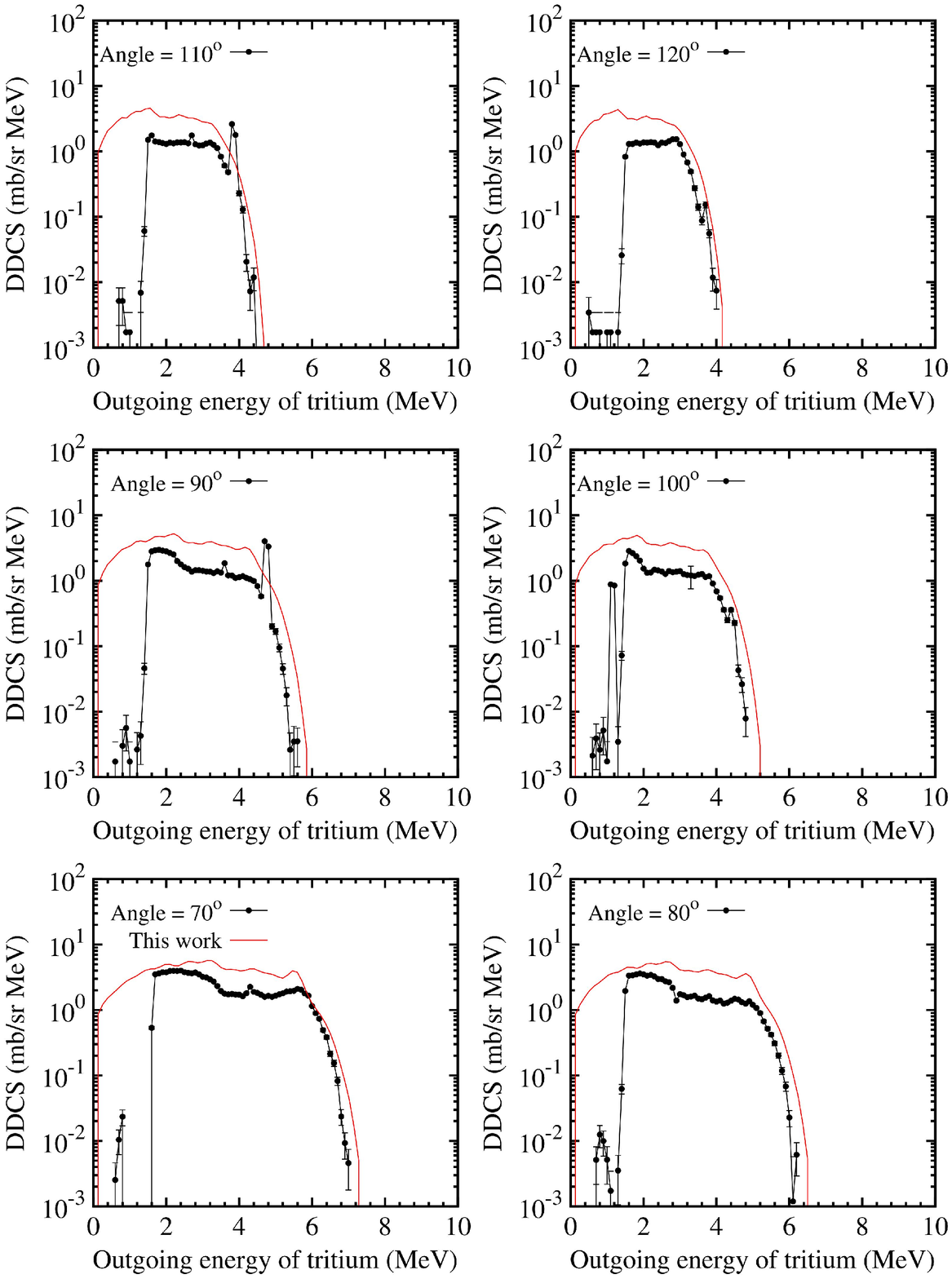}
\caption{(Color online) The same as Fig. \ref{Fig7}, but for different outgoing angles as labeld in this figure.}\label{Fig8}
\end{figure}

\begin{figure}
\centering
\includegraphics[width=15cm,angle=0]{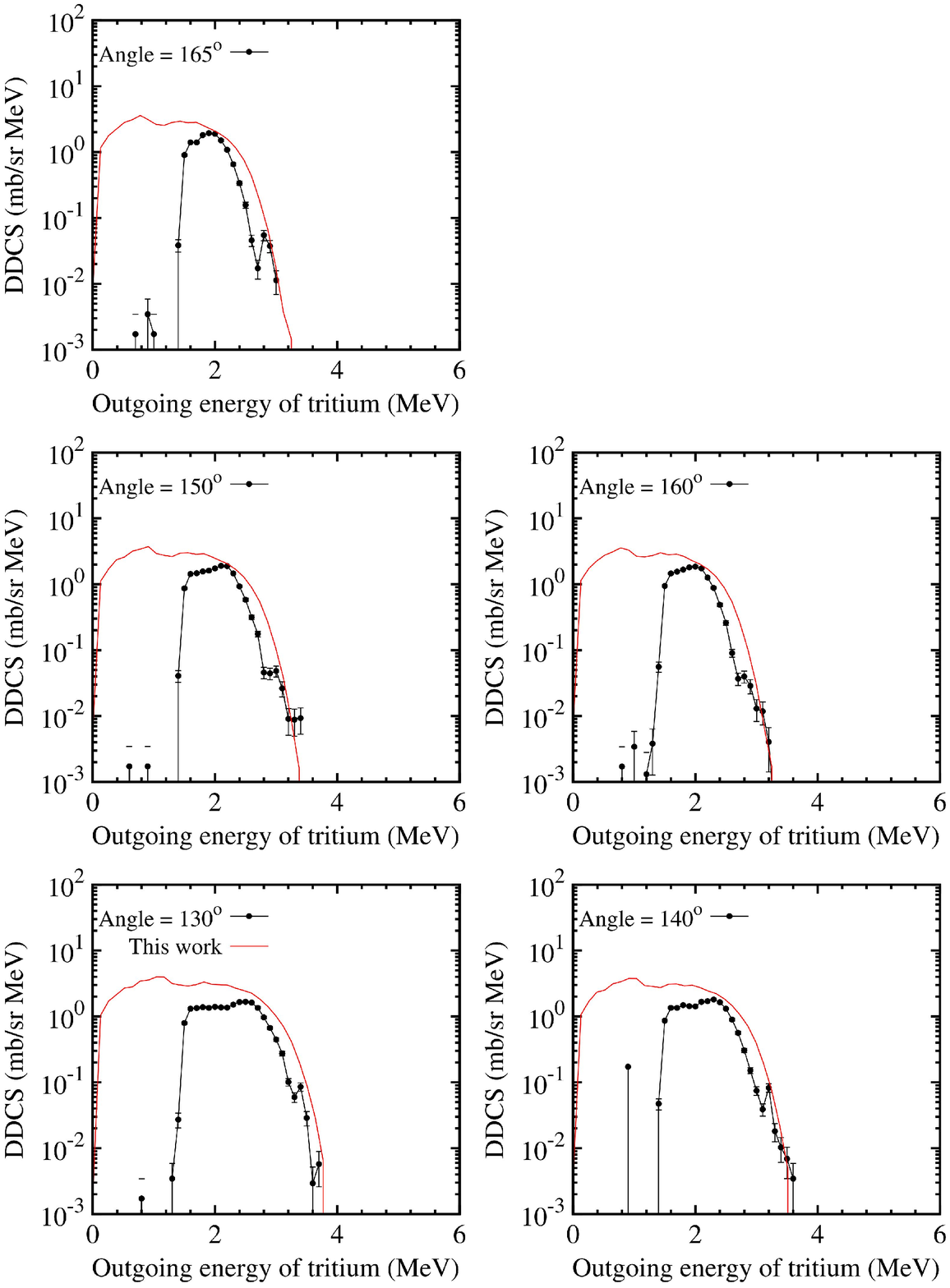}
\caption{(Color online)  The same as Fig. \ref{Fig7}, but for different outgoing angles as labeld in this figure.}\label{Fig9}
\end{figure}

\begin{figure}
\centering
\includegraphics[width=20cm,angle=0]{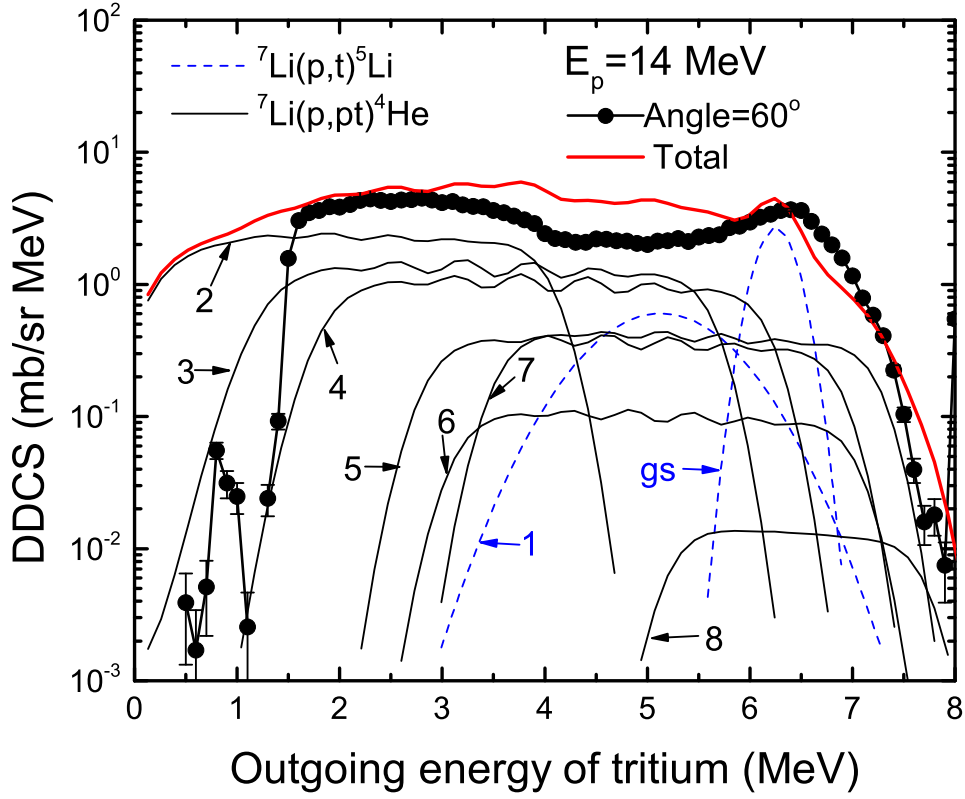}
\caption{\label{Fig_Par_T1}(Color online) The same as Fig. \ref{Fig_Par_P1}, but for outgoing triton. The blue dashed lines denote the partial spectra of first outgoing triton from the compound nucleus to the ground state and the first excited energy level of $^5$Li for $^7$Li$(p, t)^5$Li reaction channel. The black solid lines denote the contributions from second to eighth excited energy levels of $^7$Li to the ground state of $^4$He for $^7$Li$(p, pt)^4$He reaction channel. }\label{Fig15}
\end{figure}

\begin{figure}
\centering
\includegraphics[width=20cm,angle=0]{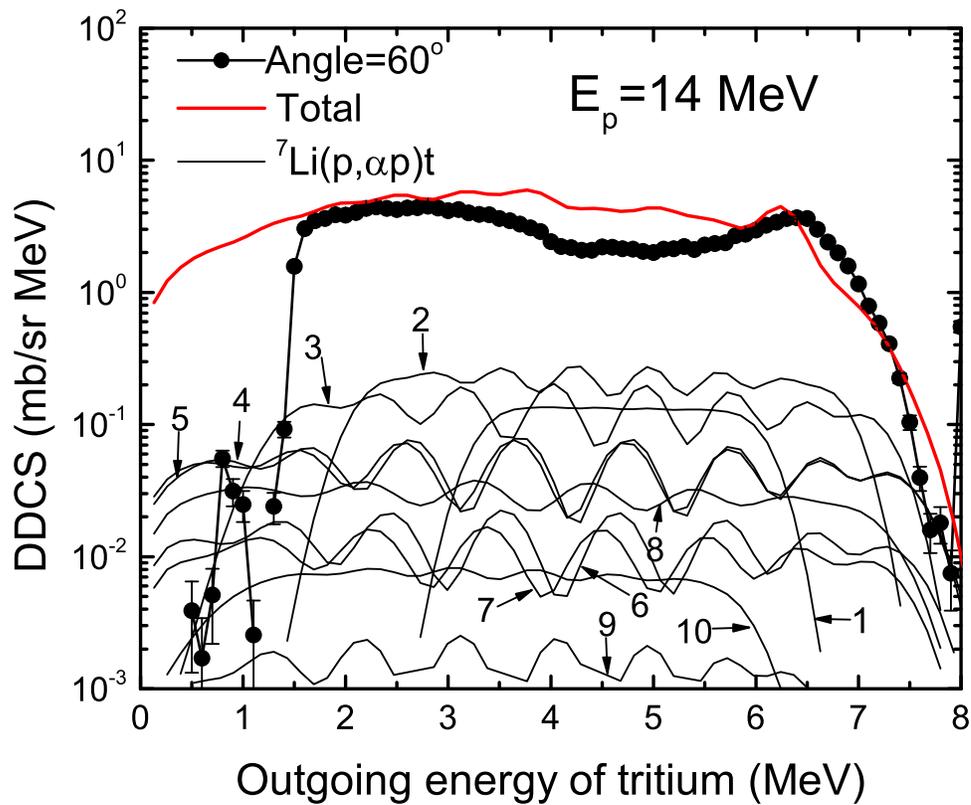}
\caption{\label{Fig_Par_T2}(Color online)  The same as Fig. \ref{Fig_Par_T1}, but the black solid lines denote the partial spectra of secondary outgoing triton from the first to 10th excited energy levels of $^4$He, which can break up into $p$ and $t$.}\label{Fig16}
\end{figure}

The calculated double differential cross sections of total outgoing triton for $p + ^7$Li reaction at $14$ MeV are compared with the experimental data with outgoing angles of $10^\circ, 20^\circ, 30^\circ, 40^\circ, 50^\circ, 60^\circ, 70^\circ, 80^\circ, 90^\circ, 100^\circ, 110^\circ, 120^\circ, 130^\circ, 140^\circ, 150^\circ, 160^\circ$ and $165^\circ$, as shown in Figs. \ref{Fig7} - \ref{Fig9}. The black points denote the experimental data derived from Ref. \cite{N. Koori1991}, and the red solid lines denote the calculated total double-differential cross sections of outgoing triton. One can see that the calculated results agree well with the measurements. Fig. \ref{Fig_Par_T1} shows the partial double differential cross sections of outgoing triton from reaction channels $^7$Li($p,t$)$^5$Li and $^7$Li($p,pt$)$^4$He with outgoing angle $60^\circ$ at $E_p = 14$ MeV in LS.
In Fig. \ref{Fig_Par_T1}, the blue dashed lines denote the partial spectra of the first outgoing triton from the compound nucleus to the ground state and the first excited energy level of $^5$Li for $^7$Li$(p, t)^5$Li reaction channel. The black solid lines denote the contributions of secondary outgoing triton from second to eighth excited energy levels of $^7$Li to the ground state of $^4$He for $^7$Li$(p, pt)^4$He reaction channel. Fig. \ref{Fig_Par_T2} shows the partial double differential cross sections of secondary outgoing triton from reaction channel $^7$Li($p,\alpha p$)$t$ with outgoing angle $60^\circ$ at $E_p = 14$ MeV in LS. In Fig. \ref{Fig_Par_T2}, the black solid lines denote the partial spectra of secondary outgoing triton from the first to 10th excited energy levels of $^4$He, which can break up into $p$ and $t$. As shown in Figs. \ref{Fig_Par_P3} and \ref{Fig_Par_T2}, there are some wave-form partial spectra because of the too small Gaussian expansion coefficients no more than.

\section{SUMMARY AND CONCLUSION}
\label{sect4}

Based on the unified Hauser-Feshbach and exciton model \cite{Zhang1993}, which can describe the nuclear reaction emission processes between the discrete levels with energy, angular momentum, and parity conservation, STLN has been applied successfully to calculate the double differential cross sections of outgoing neutrons for neutron and proton induced reactions with the $1p$-shell nuclei involved. In this paper, the STLN has been improved to calculate the double differential cross sections of outgoing neutron, proton, deuteron, triton, $^3$He, and alpha particle for proton induced $^7$Li nuclear reaction. The calculated results very well reproduce the existed measurements of outgoing proton, deuteron and triton. The model calculation for $p + ^7$Li reaction indicates that the pre-equilibrium emission process is the dominant reaction mechanism in 1$p$-shell light nucleus reactions. Due to the light mass of $1p$-shell nuclei, the recoiling effects in various emission processes are strictly taken into account. And the calculated results indicate that the double differential cross sections of outgoing particles are very sensitive to energy, spin and parity of the discrete energy levels both for the target nucleus and the corresponding residual nuclei. Furthermore, the complete nuclear reaction data with ENDF-6 format for $p + ^7$Li can be obtained by PUNF code on the basis of STLN.

\textbf{Acknowledgements}

This work is supported by Natural Science Foundation of Guangxi (No. 2019GXNSFDA185011); National Natural Science Foundation of China (No. 11465005); and Foundation of Guangxi innovative team and distinguished scholar in institutions of higher education (No. 2017GXNSFGA198001).

\end{document}